\documentclass[12pt]{article}

\usepackage[utf8]{inputenc}
\usepackage[T1]{fontenc}
\usepackage{lmodern}
\usepackage{hyperref}
\usepackage{enumitem}
\usepackage{verbatim}
\usepackage{url}
\usepackage{tikz}
\usetikzlibrary{positioning, arrows.meta}
\usepackage{float}

\title{Transforming Cyber Defense: Harnessing Agentic and Frontier AI for Proactive, Ethical Threat Intelligence}
\author{
  Dr. Krti Tallam \\
  EECS, University of California at Berkeley \\
  \texttt{ktallam@berkeley.edu}
}

\begin{document}

\maketitle

\begin{abstract}
In an era marked by unprecedented digital complexity, the cybersecurity landscape is evolving at a breakneck pace, challenging traditional defense paradigms. Advanced Persistent Threats (APTs) reveal inherent vulnerabilities in conventional security measures and underscore the urgent need for continuous, adaptive, and proactive strategies that seamlessly integrate human insight with cutting-edge AI technologies. This manuscript explores how the convergence of agentic AI and Frontier AI is transforming cybersecurity by reimagining frameworks such as the cyber kill chain, enhancing threat intelligence processes, and embedding robust ethical governance within automated response systems. Drawing on real-world data and forward-looking perspectives, we examine the roles of real-time monitoring, automated incident response, and perpetual learning in forging a resilient, dynamic defense ecosystem. Our vision is to harmonize technological innovation with unwavering ethical oversight—ensuring that future AI-driven security solutions uphold core human values of fairness, transparency, and accountability while effectively countering emerging cyber threats.
\end{abstract}

\clearpage
\section{Table of Contents}
\begin{enumerate}
  \item Introduction
  \item The Evolving Threat Landscape
    \begin{enumerate}[label*=\arabic*.]
      \item What Are Advanced Persistent Threats (APTs)?
      \item The Cyber Kill Chain and Its Limitations
    \end{enumerate}
  \item Agentic AI and Frontier AI in Cybersecurity
    \begin{enumerate}[label*=\arabic*.]
      \item Proactive Sensing, Continuous Learning, Contextual Analysis, and Responsive Modulation
      \item Rethinking the Attack Lifecycle: The Adaptive Engagement Paradigm
    \end{enumerate}
  \item Building a Cyber Threat Intelligence (CTI) Program
    \begin{enumerate}[label*=\arabic*.]
      \item Evaluating the Threat Landscape
      \item Requirements Analysis for Cyber Threat Intelligence
      \item Key Elements of a Cyber Threat Intelligence Program
    \end{enumerate}
  \item Threat Intelligence Feeds and Sources in the Era of Frontier AI
    \begin{enumerate}[label*=\arabic*.]
      \item Open Source Intelligence (OSINT)
      \item Cyber Counterintelligence (CCI)
      \item Indicators of Compromise (IoCs)
      \item Malware Analysis
    \end{enumerate}
  \item Ethical, Transparent, and Human-Centric AI Security
  \item Conclusion: A New Paradigm for Cyber Defense
  \item References
\end{enumerate}

\clearpage
\section{Introduction}
In today’s hyper-connected world, organizations face a constantly shifting array of cyber threats—ranging from brute-force malware attacks \cite{aiwithktThreats2025} to stealthy Advanced Persistent Threats (APTs) \cite{verizon2022}. Traditional security measures, such as firewalls and signature-based intrusion detection systems, are increasingly inadequate against adversaries who exploit zero-day vulnerabilities \cite{symantec2021}, \cite{aiwithktSOC2025}, employ sophisticated social engineering tactics \cite{hadnagy2018}, and maintain long-term, covert footholds within networks \cite{fireeye2020}. The financial and reputational stakes are enormous: according to a recent IBM study, the global average cost of a data breach has reached a record high, underscoring the urgent need for more robust defenses \cite{ibm2022}.

Concurrently, \textbf{Agentic AI} and \textbf{Frontier AI} are poised to transform how we think about threat intelligence, defense frameworks, and automated response \cite{aiwithkt2025apt}, \cite{aiwithkt2025cti}. These AI systems are capable of near-autonomous operation, continuous learning, and context-sensitive interventions that match or outpace the speed and sophistication of adversarial actors \cite{mitre2020}. Beyond merely enhancing existing tools, advanced AI-driven solutions promise a paradigm shift in the detection and mitigation of novel attack vectors—particularly those associated with APTs, where stealth and persistence are defining characteristics. However, their deployment also raises critical questions about ethical oversight, transparency, and alignment with human values \cite{floridi2019}. Recent guidelines from international bodies such as the European Commission emphasize the need for \emph{trustworthy} AI, highlighting principles of accountability, privacy, and fairness \cite{eu2020}.

In parallel, the cybersecurity landscape continues to grow in complexity. Rapid digital transformation, accelerated by global trends such as remote work, cloud adoption, and the proliferation of IoT devices, expands the potential attack surface for both traditional and AI-supplemented threats \cite{cybersecurityventures2021}. Security leaders are thus tasked with balancing innovation and agility against the risk of sophisticated cyberattacks. Human analysts remain indispensable for high-level strategic decisions, yet the sheer volume of threats necessitates AI systems that can autonomously handle routine tasks, filter false positives, and flag emergent anomalies \cite{gartner2021}, \cite{aiwithktSOC2025}.

This manuscript merges the technical, operational, and philosophical dimensions of cybersecurity in an age defined by AI-driven threats and defenses. By drawing upon the multi-tiered agentic AI approach, advanced threat intelligence processes, and the reimagined cyber kill chain, we aim to lay the groundwork for a more proactive, resilient, and ethically guided cybersecurity ecosystem. Ultimately, our goal is not merely to showcase new tools but to propose a holistic framework where cutting-edge AI and human expertise coexist, fostering security practices that are at once \emph{effective} and \emph{ethically grounded}.

\section{The Evolving Threat Landscape}
\subsection{What Are Advanced Persistent Threats (APTs)?}

Advanced Persistent Threats (APTs) represent a particularly insidious class of cyber threat. Unlike opportunistic attacks that aim for quick wins or widespread disruptions, APTs are characterized by long-term, strategic campaigns \cite{crowdstrike2022}. The goal is not simply to infiltrate an organization but to remain embedded for as long as possible, gathering intelligence and exfiltrating sensitive data over extended periods.

\noindent A typical APT campaign involves several carefully orchestrated stages:
\begin{enumerate}
  \item \textbf{Infiltrate}: Attackers breach the network perimeter through zero-day exploits or advanced social engineering. They often use spear-phishing emails, watering-hole attacks, or compromised third-party software updates to gain an initial foothold \cite{mandiant2021}.
  \item \textbf{Establish a Foothold}: Once inside, they deploy malware or other covert methods to maintain ongoing access. This may involve installing backdoors, creating hidden user accounts, or leveraging legitimate administrative tools to blend in with normal network traffic \cite{cisa2020}.
  \item \textbf{Explore}: Attackers systematically map the organization’s digital terrain, seeking high-value targets such as sensitive intellectual property or financial data. They often perform lateral movement—gaining escalated privileges and pivoting across systems to identify key assets \cite{lockheed2015}.
  \item \textbf{Exfiltrate}: Data is siphoned off gradually—often in small increments to evade detection—while the adversary remains embedded to gather more intelligence over time. In some cases, attackers may also alter or destroy data, adding a destructive element to the threat \cite{carnegieAPT2019}.
\end{enumerate}

Because APTs often operate stealthily for extended durations, they challenge traditional detection methods reliant on signature-based scanning or reactive incident response \cite{verizon2022}. Their adaptive, persistent nature calls for an equally dynamic and proactive approach to threat intelligence and response. In practice, this means continuously monitoring network traffic for anomalies, leveraging threat hunting techniques to identify suspicious lateral movement, and maintaining robust endpoint visibility. It also entails an organizational shift toward a zero-trust security model, where no user or system is inherently trusted \cite{nist2021}.

Notably, high-profile APT incidents have been linked to nation-state actors or well-funded criminal syndicates with specific strategic objectives, such as economic espionage or geopolitical advantage \cite{fireeye2020}. These campaigns can last months—or even years—before detection, underscoring the need for advanced analytics, machine learning-based anomaly detection, and human expertise working in tandem to thwart sophisticated adversaries. In many respects, APTs have become the defining challenge of modern cybersecurity: persistent, covert, and capable of causing severe damage if left unchecked.

\subsection{The Cyber Kill Chain and Its Limitations}

Historically, the \textbf{cyber kill chain} has served as a linear model describing the stages of an attack—from initial reconnaissance to data exfiltration. While useful for dissecting an attacker’s methodology, the kill chain can inadvertently encourage reactive defenses. As soon as attackers evolve or move laterally, defenders risk being a step behind.

Furthermore, the introduction of \textbf{agentic AI} into cybersecurity defense suggests that a static, militaristic kill chain model may no longer suffice. Agentic AI systems can operate continuously—monitoring, learning, and adapting in real time—thus demanding a more fluid and interconnected framework for defense.

\section{Agentic AI and Frontier AI in Cybersecurity}

Agentic AI refers to autonomous or near-autonomous systems capable of making end-to-end decisions without extensive human oversight. Such systems integrate advanced machine learning algorithms with real-time data streams to rapidly detect and respond to emerging threats. In contrast, \textbf{Frontier AI}—highly advanced generative models with the ability to reason, adapt, and self-learn—pushes these capabilities even further. Frontier AI is not only capable of processing vast amounts of data but can also generate novel insights and predictive models, making it an invaluable tool in combating sophisticated cyber adversaries.

The promise of these AI systems lies in their potential to transform cybersecurity from a reactive practice into a proactive, continuously adaptive defense mechanism. By automating routine threat detection and response tasks, these systems allow human analysts to focus on higher-level strategic decisions, ultimately enhancing an organization’s overall security posture.

\subsection{Proactive Sensing, Continuous Learning, Contextual Analysis, and Responsive Modulation}

A robust AI-driven cybersecurity framework relies on a continuous feedback loop comprising four key components. The following conceptual diagram illustrates this cyclical and adaptive approach, as discussed in \cite{aiwithkt2025apt} and \cite{aiwithkt2025cti}.

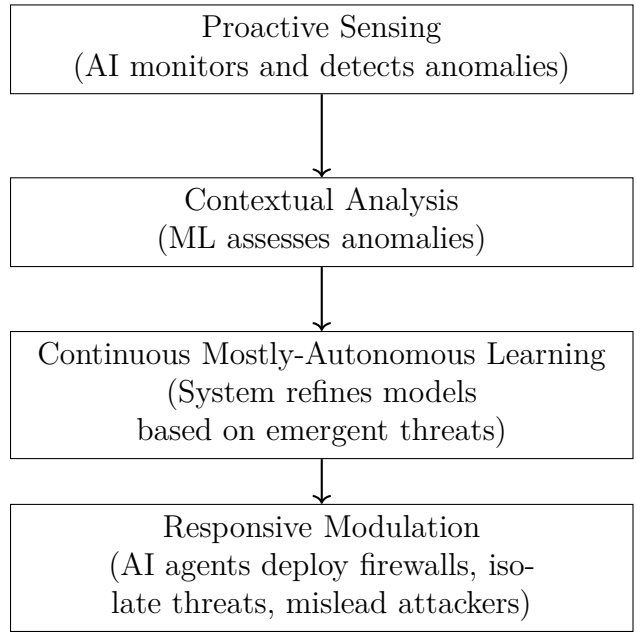
\begin{figure}[H]
\centering
\begin{tikzpicture}[
    node distance=2.3cm, 
    auto, 
    every node/.style={rectangle, draw, align=center, text width=8cm}
]
    \node (A) {Proactive Sensing\\(AI monitors and detects anomalies)};
    \node (B) [below of=A] {Contextual Analysis\\(ML assesses anomalies)};
    \node (C) [below of=B] {Continuous Mostly-Autonomous Learning\\(System refines models based on emergent threats)};
    \node (D) [below of=C] {Responsive Modulation\\(AI agents deploy firewalls, isolate threats, mislead attackers)};
    
    \draw[->, thick] (A) -- (B);
    \draw[->, thick] (B) -- (C);
    \draw[->, thick] (C) -- (D);
\end{tikzpicture}
\caption{Conceptual diagram of AI-driven cybersecurity workflow: From Proactive Sensing to Responsive Modulation.}
\label{fig:ai_workflow}
\end{figure}

\begin{enumerate}
  \item \textbf{Proactive Sensing}:  
    AI agents are deployed to continuously monitor network traffic, system logs, and user behavior. By establishing a baseline of “normal” activity, these agents can detect even minor deviations indicative of potential threats. For example, a sudden spike in outbound data or an unusual login time might trigger further investigation. Proactive sensing is critical because it enables early detection, often before an attacker can achieve full network penetration.
  
  \item \textbf{Contextual Analysis}:  
    Once an anomaly is detected, machine learning algorithms analyze the context to determine the severity and likelihood of malicious intent. This step involves correlating various data points—such as IP addresses, behavioral patterns, and historical threat data—to differentiate between benign anomalies and genuine threats. Advanced analytics can, for instance, distinguish between a legitimate software update and a covert data exfiltration attempt. The system may use clustering techniques, anomaly detection models, and even natural language processing for analyzing logs and communications.

  \item \textbf{Responsive Modulation}:  
    When a threat is confirmed or suspected, the system initiates automated responses. This may include isolating affected network segments, updating firewall rules, or deploying decoy systems (honeypots) to mislead attackers. Responsive modulation is designed to be both swift and proportional: while some threats require immediate isolation of critical infrastructure, others might only need temporary monitoring. The goal is to mitigate damage while preserving system functionality and minimizing disruption to legitimate operations.

  \item \textbf{Continuous Mostly-Autonomous Learning}:  
    Every incident—whether a false positive or a genuine threat—feeds back into the system’s learning pipeline. This continuous learning process refines the AI models over time, enhancing their ability to detect subtle patterns and adapt to evolving attack methodologies. As the system processes more data, it becomes increasingly proficient at predicting potential threats and adjusting its defensive strategies accordingly. This dynamic evolution is essential in an environment where adversaries continuously change their tactics.
\end{enumerate}

Together, these components form a closed-loop system that not only detects and responds to cyber threats but also learns from each encounter to improve future performance. This holistic approach ensures that the cybersecurity infrastructure remains resilient against rapidly evolving threats, while reducing the burden on human analysts and enhancing overall operational efficiency \cite{aiwithkt2025apt}.

\subsection{Rethinking the Attack Lifecycle: The Adaptive Engagement Paradigm}

In the evolving landscape of cyber threats, the traditional, linear model of the kill chain is giving way to a more fluid, adaptive framework. This new paradigm — what we call the \textbf{Adaptive Engagement Paradigm}, as discussed in \cite{aiwithkt2025apt} and \cite{aiwithkt2025cti}
 — recognizes that modern attacks are neither neatly segmented nor static. Instead, they are dynamic processes in which adversaries continuously adapt their tactics. This section outlines a holistic approach that leverages agentic AI to engage attackers in real time, disrupt their operations at multiple points, and evolve through continuous learning—all while adhering to ethical constraints.

\textbf{Dynamic Engagement Continuum}:  
Rather than relying on a rigid, step-by-step progression, this paradigm envisions a \textbf{continuous engagement} model. Here, agentic AI operates in parallel with an attacker's moves. From the moment an adversary initiates reconnaissance to the later stages of lateral movement and exfiltration, AI systems are deployed to continuously monitor, analyze, and intervene. This dynamic continuum enables the defense to shift from a purely reactive posture to one that is predictive and preemptive. However, it is essential to acknowledge the current limitations of AI—such as the risk of overfitting to historical patterns and the challenge of distinguishing between benign anomalies and novel attack vectors—which necessitates ongoing refinement and human oversight.

\textbf{Intelligence-Driven Intervention Points}:  

Within this dynamic model, the system can implement interventions at multiple critical junctures:
\begin{enumerate}
  \item \textbf{Early Warning}: At the initial signs of reconnaissance—such as unusual login patterns or abnormal network probes—AI systems can flag these activities. By correlating these early indicators with known threat intelligence, the system can alert security teams and automatically adjust monitoring thresholds.
  \item \textbf{Mid-Attack Containment}: During the lateral movement phase, when attackers attempt to escalate privileges or move laterally across the network, the system can take swift action. This might include blocking malicious IP addresses, isolating compromised nodes, or quarantining suspicious user accounts. In parallel, decoy systems such as honeypots can be activated to misdirect the adversary and collect detailed intelligence on their tactics, techniques, and procedures (TTPs).
  \item \textbf{Post-Incident Analysis}: After an attack or during periods of low activity, the system engages in thorough post-incident analysis. Here, data from the attack is used to update machine learning models, refine detection rules, and augment future responses. This stage is crucial for adapting to evolving attack methods and for closing any gaps in the defense.
\end{enumerate}

\textbf{Collaborative Defense Ecosystem}:  
Agentic AI, though powerful, does not operate in isolation. Instead, it is part of a broader, interconnected \textbf{collaborative defense ecosystem}. In this ecosystem, multiple AI agents—each specialized in areas such as endpoint protection, network monitoring, and cloud security—share real-time intelligence and work together to form a “defense web.” This collective approach not only increases the speed and accuracy of threat detection but also helps in building a unified response strategy that leverages insights from various security domains. While this futuristic model promises significant improvements in security posture, it also requires robust integration mechanisms and standardized communication protocols to be truly effective in real-world scenarios.

\textbf{Ethical and Transparent Decision-Making}:  
Despite the promise of AI-driven automation, ethical considerations must remain at the forefront. Automated responses—like blocking user accounts or scanning private data—carry the risk of overreach and unintended consequences. Therefore, \textbf{ethical guardrails} are imperative. Human oversight is essential to ensure that AI decisions are both proportionate and transparent. Mechanisms such as audit trails, explainable AI models, and periodic ethical reviews must be integrated into the system. This not only ensures compliance with privacy regulations but also builds trust among stakeholders by making the AI’s decision-making processes understandable and accountable.

In summary, the Adaptive Engagement Paradigm represents a futuristic yet grounded approach to cybersecurity. It embraces continuous, real-time engagement with threats, multi-layered intervention points, and a collaborative network of AI defenses—all while embedding ethical principles into its core operations. This paradigm acknowledges current technological limitations and the evolving nature of cyber threats, making it both a visionary and practical framework for future cybersecurity strategies.

\section{Building an AI Cyber Threat Intelligence (CTI) Program}

Building an effective AI Cyber Threat Intelligence (CTI) program is critical in today’s rapidly evolving cybersecurity landscape. It involves not only deploying advanced AI systems but also integrating human expertise and well-defined processes to create a dynamic, resilient, and actionable intelligence framework. The following subsections outline the key steps and components necessary to construct such a program.

\subsection{Evaluating the Threat Landscape}

An effective CTI program begins with a nuanced understanding of the organization’s \textbf{threat landscape}. This requires continuous monitoring and regular reassessment, ensuring that security measures remain aligned with emerging risks. Key steps include:

\begin{enumerate}
  \item \textbf{Identifying Key Threats}:  
    The first step is to map out who the adversaries are—be they cybercriminals, nation-state actors, hacktivists, or insider threats—and to understand their motivations and objectives. This involves studying historical incidents, current threat reports, and intelligence feeds. Recognizing that threats are not static but evolve over time is fundamental.
  
  \item \textbf{Assessing Security Posture}:  
    A thorough evaluation of the organization’s current security infrastructure is necessary. This means auditing existing tools, detection capabilities, incident response procedures, and overall readiness. Factors such as system architecture, data flow, and access controls need to be scrutinized to determine vulnerabilities.
  
  \item \textbf{Mapping Vulnerabilities}:  
    Once key threats and security gaps are identified, it is essential to pinpoint where these vulnerabilities exist. This includes technological shortcomings—such as outdated software or unpatched systems—as well as human-factor weaknesses, like insufficient training or social engineering risks. Regular vulnerability assessments and penetration testing can provide the insights needed to update this \textbf{living document} continuously.
\end{enumerate}

A dynamic threat landscape requires that the CTI program is not a one-time setup but a continuously updated resource that evolves with the digital ecosystem.

\subsection{Requirements Analysis for Cyber Threat Intelligence}

With the threat landscape mapped, the next step is to conduct a comprehensive requirements analysis to ensure that the CTI program is tailored to the organization's specific needs. Key areas to focus on include:

\begin{enumerate}
  \item \textbf{Defining Objectives}:  
    Clearly articulate what the CTI program is designed to achieve. Objectives may range from early detection of APTs and rapid incident response to providing strategic insights for executive decision-makers. Establishing these objectives upfront helps in prioritizing resources and aligning them with business needs.
  
  \item \textbf{Stakeholder Alignment}:  
    Reconciling the diverse needs of various stakeholders—security teams, IT, management, and even external partners—is crucial. This involves engaging in cross-functional discussions to define what success looks like for each group and ensuring that the CTI program delivers actionable intelligence to all relevant parties.
  
  \item \textbf{Establishing Engagement Rules}:  
    Secure collaboration within the CTI framework is essential. Define protocols for data sharing, including non-disclosure agreements (NDAs) and risk thresholds. These rules help to build trust and ensure that sensitive intelligence is shared appropriately and securely across the organization.
  
  \item \textbf{Prioritizing Threats}:  
    Not all threats require the same level of attention. Prioritize threats based on potential impact and likelihood. For instance, if the organization handles highly sensitive data, then APTs might be prioritized over more generic malware attacks. This targeted focus ensures that critical vulnerabilities receive the necessary resources and response mechanisms.
\end{enumerate}

A detailed requirements analysis creates a clear roadmap for the CTI program, setting the stage for the effective deployment of tools and processes.

\subsection{Key Elements of a Cyber Threat Intelligence Program}

Once objectives and requirements are defined, the CTI program should be built upon four fundamental pillars:

\begin{enumerate}
  \item \textbf{People}:  
    Skilled analysts and decision-makers are at the heart of any CTI program. These professionals interpret raw data, validate findings, and provide context to intelligence reports. Continuous training and certification programs are essential to keep their skills current with evolving threats and technologies.
  
  \item \textbf{Process}:  
    A structured, repeatable process is critical for the consistent collection, analysis, and dissemination of threat data. This includes standardized procedures for data ingestion from various sources (such as OSINT, dark web monitoring, and internal logs), as well as workflows for incident response, threat correlation, and reporting. Documented procedures ensure that the CTI program remains effective even as team members change.
  
  \item \textbf{Technology}:  
    Advanced platforms are necessary to support the CTI program. These may include Security Information and Event Management (SIEM) systems, Endpoint Detection and Response (EDR) tools, and specialized AI-driven threat intelligence solutions. The technology stack should be scalable, integrating both legacy systems and next-generation tools, and must be capable of processing large volumes of data in real time.
  
  \item \textbf{Budgeting}:  
    Adequate financial resources are crucial not only for acquiring state-of-the-art tools but also for ongoing training, process refinement, and adapting to emerging threats. A well-funded CTI program can afford regular updates and maintenance, ensuring that its capabilities remain aligned with the latest threat intelligence and industry best practices.
\end{enumerate}

These pillars must interact seamlessly to form a robust communication plan. Intelligence should be disseminated effectively to all stakeholders, with clear performance metrics and continuous improvement cycles ensuring that the program adapts to new challenges over time.

In summary, building an AI Cyber Threat Intelligence program is not just about technology—it is a holistic endeavor that integrates people, processes, technology, and funding to create a dynamic, proactive defense mechanism. By continuously evaluating the threat landscape, conducting detailed requirements analysis, and establishing a foundation built on robust pillars, organizations can ensure that their CTI program is both resilient and adaptable in the face of ever-evolving cyber threats.

\section{Threat Intelligence Feeds and Sources in the Era of Frontier AI}

A \textbf{threat intelligence feed} is a continuous stream of actionable data—such as Indicators of Compromise (IoCs), malicious domains, and suspicious IP addresses—sourced from a variety of channels. These feeds empower security teams to anticipate, detect, and respond to emerging threats with increased precision. However, as cyber threats become more sophisticated, so too must the methods for gathering and interpreting this intelligence. The advent of \textbf{Frontier AI} brings transformative potential to this space by dramatically increasing the speed, scale, and depth of data analysis, while also introducing new complexities in data verification and contextualization. In this section, we explore the primary sources of threat intelligence and assess the impact of Frontier AI on each domain.

\subsection{Open Source Intelligence (OSINT)}

Traditionally, OSINT has served as a cornerstone for gathering publicly available information to inform threat intelligence. Typical sources include:
\begin{itemize}
  \item \textbf{Search engines and web services:} These platforms allow analysts to extract large volumes of data from the internet, ranging from news articles to technical blogs.
  \item \textbf{Website footprinting:} Analyzing website structures, domain registration details, and server configurations can reveal vulnerabilities and potential attack vectors.
  \item \textbf{Email and DNS interrogation:} Techniques such as analyzing email headers and DNS records help in identifying spoofed communications and unusual domain resolutions.
  \item \textbf{WHOIS lookups:} These provide essential information about domain ownership and history, which can be critical when tracking malicious actors.
  \item \textbf{Automated tools and scripts:} Custom scripts and frameworks aggregate data from multiple public sources, enabling continuous monitoring and trend analysis.
\end{itemize}

\textbf{Frontier AI Impact:}  
Frontier AI augments OSINT capabilities in several significant ways:
\begin{itemize}
  \item \textbf{AI-Powered OSINT Aggregators:} Leveraging advanced natural language processing and machine learning algorithms, these aggregators can rapidly scan billions of public data sources, detect subtle anomalies, and cross-reference disparate datasets to uncover hidden threat patterns.
  \item \textbf{Deepfake and AI-Generated Misinformation:} The rise of synthetic media challenges traditional OSINT methods. Frontier AI can help develop robust verification frameworks to differentiate authentic content from manipulated or fabricated information.
  \item \textbf{Real-Time Adaptive Reconnaissance:} Attackers are increasingly using AI to perform adaptive reconnaissance that evolves in real time. Defensive systems must, therefore, employ equally dynamic AI tools that can continuously learn and adapt to changing data sources and methodologies.
\end{itemize}

This bolstered OSINT capability allows organizations to build richer, more context-aware threat profiles while continuously validating the reliability of publicly available data.

\subsection{Cyber Counterintelligence (CCI)}

Cyber Counterintelligence (CCI) focuses on gathering intelligence on adversaries by directly monitoring their activities. Traditional CCI methods include:
\begin{itemize}
  \item \textbf{Passive DNS Monitoring:} Observing DNS queries and responses to detect patterns that may indicate domain flux or other malicious activities.
  \item \textbf{Honeypots:} Deployed to attract and analyze attacker behavior in a controlled environment, honeypots serve as decoys that reveal adversary tools and techniques.
  \item \textbf{Infrastructure Pivoting:} By infiltrating adversary-controlled systems or networks, defenders can gather valuable intelligence on attack methodologies and command-and-control structures.
  \item \textbf{Malware Sinkholes:} These systems capture and analyze malware traffic, helping to identify new variants and their propagation mechanisms.
  \item \textbf{YARA Rules:} Used to classify and detect malware by matching patterns in code, YARA rules remain a staple in malware classification.
\end{itemize}

\textbf{Frontier AI Impact:}  
Frontier AI brings several advancements to the realm of CCI:
\begin{itemize}
  \item \textbf{AI-Driven Attack Simulation:} Advanced AI models can simulate complex attack scenarios, enabling defenders to anticipate adversary behavior and test the effectiveness of their countermeasures in a virtual environment.
  \item \textbf{Autonomous Cyber-Adversaries:} There is an emerging risk that threat actors will deploy self-learning AI agents capable of launching autonomous attacks. This necessitates the development of equally autonomous defensive agents that can counteract such threats.
  \item \textbf{AI-Augmented Honeypots:} Next-generation honeypots are being designed to adapt in real time, modifying their appearance and behavior to more convincingly mimic real systems, thereby gathering more sophisticated intelligence on attacker tactics.
\end{itemize}

By integrating these advanced techniques, organizations can gain deeper insights into adversary behavior and more effectively disrupt their operations before significant damage occurs.

\subsection{Indicators of Compromise (IoCs)}

Indicators of Compromise (IoCs) are specific artifacts that signal a potential security breach. These may include:
\begin{itemize}
  \item \textbf{Network Traffic Analysis:} Detecting anomalies in data flow, such as unusual outbound connections or unexpected data volumes.
  \item \textbf{Host-Based Artifacts:} Logs, file hashes, and system events that can reveal unauthorized activities.
  \item \textbf{Threat Intelligence Sharing Platforms:} Collaborative systems that allow organizations to exchange IoCs and other threat data, enhancing collective security.
\end{itemize}

\textbf{Frontier AI Impact:}  
The application of Frontier AI in IoC management introduces both opportunities and challenges:
\begin{itemize}
  \item \textbf{AI-Augmented IoC Identification:} Frontier AI systems can process vast quantities of data to identify subtle, previously unnoticed patterns, significantly enhancing early detection capabilities.
  \item \textbf{Automated IoC Obfuscation by Attackers:} Adversaries increasingly use AI to modify and obfuscate IoCs, creating polymorphic malware and continuously mutating code. Defensive systems must evolve to detect these rapidly changing signatures.
  \item \textbf{AI-Generated Threat Intelligence Reports:} By automating the analysis and classification of IoCs, Frontier AI can generate comprehensive threat reports in real time, enabling organizations to adjust their defensive measures proactively.
\end{itemize}

These advancements help ensure that IoCs remain reliable and actionable, even as adversaries employ increasingly sophisticated techniques to hide their activities.

\subsection{Malware Analysis}

Malware analysis is the process of dissecting malicious software to understand its structure, behavior, and potential impact. Traditional malware analysis involves:
\begin{itemize}
  \item \textbf{Dissecting Malware Samples:} Reverse-engineering malware to reveal its code structure and operational logic.
  \item \textbf{Identifying Origins and Capabilities:} Determining the source of malware and assessing its potential to cause harm.
  \item \textbf{Observing Malware Interactions:} Monitoring how malware interacts with system resources and networks to understand its propagation and infection mechanisms.
\end{itemize}

\textbf{Frontier AI Impact:}  
Frontier AI is poised to significantly transform malware analysis:
\begin{itemize}
  \item \textbf{AI-Generated Malware:} Adversaries may leverage Frontier AI to design novel malware strains that are optimized for stealth, evasion, and rapid adaptation, presenting a formidable challenge to traditional analysis techniques.
  \item \textbf{Automated Exploit Generation (AEG):} Advanced AI can expedite the creation of exploits by automatically generating code targeting newly discovered vulnerabilities. This reduces the window between vulnerability disclosure and exploitation in the wild.
  \item \textbf{AI-Driven Malware Detection:} By employing large-scale pattern recognition and real-time analysis, Frontier AI systems can identify emerging threats more quickly and accurately than conventional signature-based methods.
\end{itemize}

In addition to these technical capabilities, the integration of Frontier AI into malware analysis promises to escalate the speed and accuracy of threat identification, enabling more proactive defensive measures. However, this increased sophistication also demands greater vigilance and robust validation methods to ensure that automated insights are both reliable and actionable.

\noindent In summary, the era of Frontier AI is transforming threat intelligence feeds by enhancing the volume, variety, and granularity of data available to security teams. While these advances offer unprecedented opportunities for early detection and proactive defense, they also require organizations to address new challenges related to data authenticity, algorithmic complexity, and the balance between automation and human oversight. Ultimately, the goal is to create a dynamic, continually evolving ecosystem where human expertise and machine intelligence work in tandem to safeguard digital assets in an increasingly complex threat landscape.

\section{Ethical, Transparent, and Human-Centric AI Security}

As AI systems become increasingly autonomous in the realm of cybersecurity, fundamental questions of \textbf{agency, accountability, and ethics} must be addressed. While advanced AI technologies promise amplified detection and faster responses, they also raise critical concerns about how decisions are made and who is ultimately responsible. In this context, several key principles guide the development and deployment of AI security systems:

\begin{itemize}
  \item \textbf{Transparency}:  
    For AI-driven security decisions—especially those impacting user privacy, critical infrastructure, or sensitive data—it is imperative that the decision-making process is explainable. Transparency involves clear documentation of algorithms, data sources, and reasoning behind automated actions. This ensures that stakeholders can understand, audit, and trust the system's outputs, reducing the risk of unexpected or biased decisions.
    
  \item \textbf{Accountability}:  
    As responsibility shifts from human operators to autonomous systems, mechanisms must be in place to trace AI-driven actions back to accountable individuals or governance structures. This includes implementing audit trails, logging all critical decisions, and establishing clear policies regarding liability and oversight. Accountability not only fosters trust but also provides a framework for continuous improvement by learning from mistakes.
    
  \item \textbf{Human Oversight}:  
    Although AI can process data and make decisions at scales beyond human capability, it cannot fully replace the nuanced judgment of experienced cybersecurity professionals. Human oversight remains essential, particularly for handling edge cases, interpreting ambiguous situations, and making ethical determinations that require empathy and contextual understanding. By integrating a human-in-the-loop model, organizations can ensure that automated processes complement rather than replace human expertise.
    
  \item \textbf{Fairness and Inclusivity}:  
    AI systems must be designed to avoid perpetuating biases and inequities. This requires diverse training data, inclusive design practices, and regular reviews of system outputs to identify and mitigate bias. Fairness is crucial not only from an ethical standpoint but also to ensure that security measures do not inadvertently discriminate against certain groups or create unintended vulnerabilities.
    
  \item \textbf{Robustness and Safety}:  
    Given the potential for adversaries to exploit vulnerabilities in AI systems, robust safety mechanisms must be embedded within AI architectures. This includes fail-safe protocols, redundancy, and continuous testing under real-world conditions. A focus on safety ensures that AI systems remain reliable even under attack, reducing the likelihood of catastrophic failures.
\end{itemize}

In practice, these principles require the development of an integrated ethical framework that combines technical safeguards with policy and governance measures. For instance, organizations should establish interdisciplinary ethics committees that include experts in cybersecurity, law, and social sciences to oversee AI implementations. Regular external audits and adherence to international standards—such as those proposed by the European Commission for trustworthy AI—can further reinforce ethical practices.

Moreover, fostering a \textbf{collaborative model} where AI augments human defenders is essential. Rather than pitting machine intelligence against human judgment, the goal is to create systems where both operate in synergy. AI can handle repetitive tasks, analyze vast data sets, and flag potential threats, while human operators provide critical context, moral reasoning, and strategic oversight. This partnership not only improves the efficiency and accuracy of threat detection and response but also ensures that ethical considerations are embedded in every level of operation.

Ultimately, building ethical, transparent, and human-centric AI security systems is not a one-time effort but an ongoing process. As technologies evolve and threats become more sophisticated, continuous refinement of ethical guidelines, accountability mechanisms, and oversight protocols will be necessary. In doing so, organizations can harness the full potential of AI to protect digital assets while upholding the values and rights fundamental to a just and secure society.

\section{A New Paradigm for Cyber Defense}
The integration of \textbf{agentic AI} and \textbf{Frontier AI} into cybersecurity represents not just an incremental improvement in technology, but a \textbf{transformative shift} in how we understand and implement digital defense. This shift compels us to revisit and fundamentally reimagine our assumptions about intelligence, responsibility, and ethical governance. Traditional linear models, such as the cyber kill chain, are no longer sufficient in an environment where Advanced Persistent Threats (APTs) evolve in real time and adversaries continually adapt their strategies.

In this new paradigm, defense is conceived as a \textbf{dynamic, continuously engaged} process. Rather than reacting to threats only after they have manifested, organizations are now positioned to anticipate and counteract malicious activities at every stage of an attack. By weaving together \textbf{proactive sensing}, \textbf{contextual analysis}, \textbf{responsive modulation}, and \textbf{continuous learning}, a layered and adaptive defense is established—one that is capable of intercepting threats before they fully materialize.

\textbf{Proactive sensing} ensures that every anomaly in network traffic, user behavior, or system performance is monitored in real time. This continuous vigilance enables the early detection of potential threats, providing the opportunity to analyze and neutralize risks before they escalate. Through \textbf{contextual analysis}, raw data is transformed into actionable intelligence; advanced algorithms assess the significance of anomalies, correlating them with historical patterns and emerging trends to accurately identify real threats amid a sea of noise.

\textbf{Responsive modulation} then takes center stage by dynamically deploying countermeasures. This may involve isolating compromised segments, updating firewall policies, or even triggering preemptive alerts to security teams. By acting swiftly, the system can contain and neutralize threats without waiting for full-scale breaches. Finally, \textbf{continuous learning} is embedded into the process, enabling the defense mechanism to evolve. Every incident—whether a false alarm or a genuine attack—feeds into the system's knowledge base, allowing for iterative improvements and the refinement of predictive models.

However, the transformative potential of this paradigm is inseparable from its ethical dimensions. The convergence of \textbf{technology and human judgment} is central to ensuring that these advanced systems operate not only efficiently but also in alignment with societal values. Transparency in algorithmic decision-making, robust accountability frameworks, and ongoing human oversight are critical to ensure that automated processes do not compromise privacy or civil liberties. This ethical commitment ensures that, while the technology adapts at a breakneck pace, the principles that define our humanity—fairness, responsibility, and trust—remain intact.

This paradigm is a call to action for cybersecurity professionals, AI researchers, policymakers, and stakeholders across industries. Collaboration is essential to develop integrated systems that balance cutting-edge innovation with ethical governance. By fostering environments where human expertise and machine intelligence work in concert, we can build a security posture that is agile, anticipatory, and resilient against emerging threats. In doing so, we not only protect digital assets but also ensure that our digital future remains \textbf{secure} and \textbf{aligned} with the core principles of transparency, accountability, and human dignity.

\bibliographystyle{plain}
\bibliography{references}

\end{document}